\documentclass[a4paper,11pt]{article}%
\usepackage{jheppub}
\usepackage[T1]{fontenc}%

\usepackage{amsmath}%
\usepackage{amsfonts}%
\usepackage{amssymb}%
\usepackage{graphicx}
\providecommand{\U}[1]{\protect\rule{.1in}{.1in}}
\pdfoutput=1

\def\eps{\varepsilon}

\def \RR {{\mathbb R}}

\def \eins {{\mathbf 1}}

\def\ol{\overline}

\newcommand{\later}{\succcurlyeq}
\def\eref#1{(\ref{#1})}
\def\aref#1{App.\ \ref{#1}}
\def\sref#1{Sect. \ref{#1}}
\def\bea#1{\begin{eqnarray}\label{#1}}
\def\eea{\end{eqnarray}}
\def\ba{\begin{array}}
\def\ea{\end{array}}

\title{Gauss' Law and String-Localized Quantum
  Field Theory }
\author[a]{Jens Mund}
\affiliation[a]{Departamento de F\'isica, Universidade Federal de Juiz
  de Fora, \\ Campus Universit\'ario da UFJF, Juiz de Fora 36036-900, MG, Brasil}
\emailAdd{jens.mund@ufjf.edu.br}
\author[b]{Karl-Henning Rehren}
\affiliation[b]{Institute for Theoretical Physics,
  Georg-August-University G\"ottingen,\\ 
  Friedrich-Hund-Platz 1, 37077 G\"ottingen, Germany}
\emailAdd{krehren@gwdg.de}
\author[c]{Bert Schroer}
\affiliation[c]{Centro Brasileiro de Pesquisas F\'isicas, \\ Rua
  Dr. Xavier Sigaud 150, 22290-180 Rio de Janeiro, RJ, Brasil;
  \\Institut f\"ur Theoretische Physik, Freie Universität Berlin, \\ Arnimallee 14, 14195 Berlin, Germany}
\emailAdd{schroer@zedat.fu-berlin.de}


\abstract{\noindent The quantum Gauss Law as an
  interacting field equation is a prominent feature of QED with
  eminent impact on its algebraic and superselection
  structure. It forces charged particles to be accompanied by
  ``photon clouds'' that cannot be realized in the Fock space, and
  prevents them from having a sharp mass \cite{FMS, Bu2}.  Because it
  entails the possibility of  
  ``measurement of charges at a distance'', it is well-known to be in
  conflict with locality of charged fields in a
    Hilbert space \cite{BLOT,FPS}. We show
  how a new approach to QED advocated in \cite{MS,S,S2,MRS} that avoids
  indefinite metric and ghosts, can secure causality and
  achieve Gauss' Law along with all its nontrivial consequences. We explain why
  this is not at variance with recent results in \cite{BCRV}.
} 

\dedicated{Dedicated to Detlev Buchholz on the occasion of his 75th birthday}

\begin{document}

\maketitle

\flushbottom

\section{The quantum Gauss Law}
\label{s:GL} 
\setcounter{equation}{0} 

\subsection{Some history}
\label{s:GL1}
 
Before turning to the new results announced in the abstract, we
  want to call to mind some early developments and insights about the
infrared intricacies of QED.

The message that interactions of photons with matter require a
formulation which differs from standard scattering theory goes back to
the work of Bloch and Nordsieck \cite{BN} in 1937 and Fierz and Pauli
\cite{FP} in 1938. Decades later Yennie, Frautschi and Suura
\cite{YFS} integrated these early proposals into the covariant
renormalized perturbation theory of QED. In this way the logarithmic on-shell
infrared divergencies are encoded into the prescription for a
rotational-invariant infrared photon inclusive cross-section for the
scattering of charge-carrying particles. A closely related result of
QED perturbation theory is (what is nowadays referred to as)
Weinberg's soft photon theorem. 

As a matter of fact, apart from the rather
indirect construction by Steinmann \cite{St}, up to date these heuristic
on-shell prescriptions have not been supplemented by a direct
construction of a physical\footnote{I.e., with positive
  definite correlation functions, cf.\ \aref{a:IM}.} Dirac field in causal perturbation
theory. The present work is intended to report on substantial progress in
this direction.

Back in the early 1960s, the LSZ scattering theory (which established
the existence of in/out scattering states and the associated S-matrix as a 
consequence of the causal separability and the existence of a mass gap)
suggested that what is at stake in QED is not only the large-time LSZ
asymptotics, but the very concept of a Wigner particle with a
sharp mass-shell. At that time the only indication for such an
``infra-particle'' without a sharp mass came from a model in 1+1
spacetime dimensions by one of us \cite{S63}, that also exhibited a 
non-trivial dynamical superselection structure.

More than another decade later, Ferrari, Picasso and Strocchi 
\cite{FPS} proved that electrically charged fields cannot be pointlike 
localized\footnote{Localization is always 
  understood in the sense of causal commutation relations (vanishing
  commutator at spacelike distance). For some comments on ``causality''
  and ``observables'', see \aref{a:CO}.},
simply because local fields must commute with the charge operator,
expressed as an integral over the electric field at spacelike
infinity. Yet, it is indispensible for scattering theory 
that charged operators remain causally separable.
Fr\"ohlich, Morchio and Strocchi \cite{FMS} showed that
charged states cannot exist in the Fock representation and that
the Lorentz symmetry is broken in irreducible charged representations. 
Buchholz \cite{Bu1} refined these insights by showing that there
exists an uncountable set of superselection rules related to the
asymptotic shape of photon clouds. Lorentz symmetry connects
  sectors with different such clouds. In the present paper we shall
  reveal an intimate relation between this fact of QED and the
  superselection structure in the model of \cite{S63} -- even if 
there were no gauge fields in that model. In a follow-up paper \cite{Bu2}
Buchholz also proved that charge-carrying particles are necessarily
infra-particles. The essential input was not perturbation theory but
rather a careful formulation of the quantum Gauss Law.

In more recent times there has been a growing interest in large distance
aspects related to interactions of massless fields of helicity $h\geq
1$ with $s<1$ matter fields, as documented in the extensive monograph 
on this subject \cite{Str}. A prominent issue is the
  existence of infinitely many ``asymptotic symmetries'', whose
  associated conservation laws are responsible for the asymptotic
  superselection structure.

\subsection{Implications of the quantum Gauss Law}
\label{s:GL2}
 
We present a brief review of the quantum
Gauss Law (the measurement of the total electric charge at infinite spacelike
distance and the ensuing failure of locality of charged fields) and
its close relation to the uncountable set of ``photon cloud'' 
superselection rules. This cannot be understood in terms of the
gauge-dependent point-localized matter fields of gauge theory 
that simply evade the conflict of the quantum Gauss Law with local
commutation relations by a gauge fixing term \cite{FPS},
  cf.\ \aref{a:IM}, rather than address its
physical consequences; instead we shall demonstrate how a new approach \cite{MS,S,S2,MRS} 
implements the many infrared features of QED without the need for state spaces
with indefinite metric (Krein spaces) and ghosts to return to positive metric.
In this approach, the physical
interacting charge-carrying matter fields become ``string-localized''
via an interaction density coupling the point-localized free fields to
``string-localized potentials''.\footnote{A ``string'' is a spacelike
  or lightlike ray
  extending from a point to infinity. (In contrast, strings of String
  Theory are a classical concept that is not reflected in causal commutation
  relations.) String-localized approaches to
  QED were previously advocated by Mandelstam \cite{M} and by
  Steinmann \cite{St,St2}.\label{f:string}} 

The Gauss Law allows to measure electric charges from a distance. The flux of
the electric field across a closed surface $\sigma$ equals the total charge in
the volume $V$ enclosed by the surface:
\begin{align}
\label{lGL}
\oint_{x^{0}=t,\,\vec{x}\in\sigma}d\vec{\sigma}(\vec{x})\vec{E}(x)=\int
_{x^{0}=t,\,\vec{x}\in V}d^{3}x\,j^{0}(x).
\end{align}
It follows from the inhomogeneous Maxwell equation (the differential Gauss
Law)
\begin{align}
\label{inhmax}
\partial_{\mu}F^{\mu\nu}(x)=j^{\nu}(x).
\end{align}
In QED, the Maxwell field strength and the charge density are
interacting quantum fields coupled through (\ref{inhmax}). The right-hand side
equals $-q$ times the Dirac current, where the unit of charge $q$ is the
coupling constant of QED.

The limiting case $V\rightarrow\RR^{3}$, the global Gauss Law
\begin{align}
\label{gGL}
\lim_{R\rightarrow\infty}\oint_{x^{0}=t,\,\vec{x}\in R\cdot S^{2}}d\vec
{\sigma}(\vec{x})\vec{E}(x)=Q,
\end{align}
where the charge operator $Q$ generates the $U(1)$ symmetry of the Dirac
field, is particularly intriguing in the quantum theory: because the
global charge operator is an integral over the field strength at
spacelike infinity, and has a nontrivial commutator with the charged field, the
latter cannot commute with the field strength at spacelike distance
\cite{BLOT,FPS}.

This fact is only the first of a number of remarkable imprints of Gauss' Law
on the algebraic structure of QED and the nature of charged particles, that
are held to be characteristic of QED as a gauge theory \cite{H}.

The global charge operator and hence the asymptotic flux operator is a
multiple of $\eins$ in every irreducible representation of the algebra of
local observables, defining the charge superselection rule. In contrast, local
flux operators are dynamical quantities that are not central operators.

The global Gauss Law also requires that in charged states, the expectation
value of the flux operator through a sphere of radius $R$ has a finite limit
as $R\to\infty$. Hence, the field strength $F_{\mu\nu}(x)$ must decay in
spacelike directions like $r^{-2}$, and the asymptotic values
\begin{align}
\label{amunu}a_{\mu\nu}(x):= \lim_{\lambda\to\infty}\lambda^{2}(\Psi,F_{\mu
\nu}(\lambda x)\Psi)
\end{align}
are eigenvalues of central observables $\lim_{\lambda\to\infty}\lambda
^{2}\,F_{\mu\nu}(\lambda x)$ in the irreducible representation of the state
described by $\Psi$. Consequently, they define uncountably many ``infrared''
super\-selection rules.  Lorentz symmetry
connects different asymptotic field configurations \eref{amunu} and
can therefore not be realized in irreducible charged sectors \cite{FMS}. Buchholz has shown in
\cite{Bu2} (cf.\ \sref{s:IP}) that the condition of non-trivial limits
in (\ref{amunu}) along with a bound on the fluctuations entails that
charged one-particle states cannot be eigenvectors of the mass 
operator $M^2=P_\mu P^\mu$, hence charged particles cannot have a
sharp mass, see \sref{s:IP}. The effect can be ascribed to
the ``infrared photon clouds'' attached to charged particles whose configuration is given
by (\ref{amunu}). This infra-particle nature of electrons is
also responsible for
the breakdown of the usual methods of scattering theory, discussed in
\sref{s:GL1}. 

It should be clear from the previous, that the issues raised
  concern properties of the interacting fields themselves, and not
  just on-shell scattering amplitudes. 
The differential Gauss Law, constituting an algebraic relation between
Maxwell and Dirac fields, is of course due to the interaction. The correct
formulation of the interacting theory must be able to properly implement it.

\subsection{The string-localized formulation of QED}
\label{s:GL3} 

In covariant quantizations of the Maxwell potential with indefinite metric,
\emph{Gauss' Law does not hold} (see \aref{a:IM}). In the usual $\lambda$ gauges,
there is a ``fictitious current'' 
$j^{\mu}_{\mathrm{fict}}=-\lambda \,\partial^{\mu}(\partial A)$, that is added to
(\ref{inhmax}) as a contribution from the free field, and thus causes
the global and local Gauss Laws to fail \cite{BLOT}, 
cf.\ also \sref{s:QED} and \sref{s:IP}. 
When the Gupta-Bleuler condition is imposed, the interacting Dirac
field is no longer defined on the 
resulting physical Hilbert space, and charged states must be constructed in a
different way, e.g., by a  limit of charge separation \cite{H}. An alternative
to the Gupta-Bleuler method to deal with the indefinite metric is pursued in
Steinmann's perturbative  construction of QED \cite{St2}.

In an emerging approach \cite{MS,S,S2,MRS}, a perturbative construction of QED
is proposed that avoids indefinite metric from the outset. Emphasizing
fundamental principles of relativistic quantum theory, this approach
aims at completing the
successful Lagrangean construction of a unitary scattering matrix of
QED by the construction of its Hilbert space fields with a full control of their
causality properties.
It starts from the fact (cf.\ \aref{a:IM}), that the equation
\[
F_{\mu\nu}(x)=\partial_{\mu}A_{\nu}(x,e)- \partial_{\nu}A_{\mu}(x,e)
\]
can be solved (as a cohomological problem) by ``string-localized potentials''
\cite{MSY}
\begin{align}
\label{slok}A_{\mu}(x,e):=\int_{0}^{\infty}ds\, F_{\mu\nu}(x+se)e^{\nu},
\end{align}
where $e$ is any (spacelike) direction in $\RR^{4}$. The ray
$\RR_+\cdot e$ is referred to as ``string''
(cf.\ footnote \ref{f:string}). They are defined (as free fields) on the physical
Hilbert space of the Maxwell field strength.

The perturbative expansion (cf.\ \sref{s:BF}) of QED with the
interaction density
\begin{align}
\label{Le}L(x,e)=A_{\mu}(x,e)j^{\mu}(x) 
\end{align}
therefore proceeds on the tensor product of the free Maxwell and Dirac
Hilbert spaces. Technically, the potential and $L(x,e)$ are
distributions also in $e$ and will require a suitable smearing (that
we suppress until \sref{s:Phot}).

We show that in this approach, the inhomogeneous Maxwell equation
(\ref{inhmax}) holds (already) in first perturbative order, and along with it
the global and local Gauss Laws. We also show how the interacting Dirac field
becomes string-localized, in accord with the NoGo result of \cite{FPS}. (This
is of course due to the string-localized interaction density; but the
non-trivial part is to understand why the Dirac field is not \emph{worse} than
string-localized.)

We shall compute in the string-localized approach the asymptotic field
configuration in states $\Psi=\psi^{*}(f)\Omega$, 
generated from the vacuum $\Omega$ by the smeared interacting Dirac field
$\psi$. The remarkable feature is that --
while the observable electromagnetic field strength remains
string-independent -- the unobservable charged field becomes
string-dependent through the interaction with the string-localized
potential, and in fact localized along the string. We find in this
situation that the asymptotic electric flux in the string-dependent
state is concentrated in the direction of the string $e$. Thus,
choosing the direction $e$ (or an average over directions), one can
generate states with ``designed'' photon clouds.

A recent proposal by Duch \cite{D} constructing the interacting field strength
by using a modified causal S-matrix in (\ref{bog}) below, also cures the global Gauss
Law. Adapting old ideas of Dollard \cite{Do} and Kulish and Faddeev \cite{FK},
it defines the S-matrix by comparison with a suitable (time-)asymptotic dynamics that
takes the (space-)asymptotic photon clouds into account, corresponding to the
``dressing transformations'' of charged
states in \cite{MS1,MS2}. The localization of the interacting Dirac field is
not addressed in that approach. 

The findings presented here are not at variance with a recent result by
Buchholz et al.\ \cite{BCRV}. The authors show (in a model with static charges, to be thought
of as a limit case of QED with infinitely heavy charges) that it is
impossible, with the help of only operators belonging to the algebra of the
Maxwell field strengths, to generate states of that algebra in which the local
flux operators take non-trivial values in accord with Gauss' Law. Namely, in
the limit considered, the local flux operators are central elements of the
algebra, and no approximations with inner conjugations can change their
values. If the algebra is given (as is understood in \cite{BCRV}), an
extension of the algebra is required. This can be conveniently achieved by
potentials with ``longitudinal degrees of freedom'' that are not present in
the field strengths. The string-localized potentials as in (\ref{slok}) cannot
be used for the purpose.

In contrast, the construction of an interacting quantum field theory involves
a change of the algebra of the free fields and of its representation.
Perturbation theory achieves these changes. This is also true if the
interaction density contains only observables, such as (\ref{Le}), because
perturbation theory is not an approximation by inner unitary conjugations with
degrees of freedom present in the interaction density. This is manifest in an
external field problem (\sref{s:Ex}), where perturbation theory is exact
and the algebra remains unchanged, but the representation is manifestly
changed: the local flux operators remain central but take the values required
by the local Gauss Law.

The actual role of the string-localized potentials in the QED
interaction density is to serve as ``catalyzers'' making the interacting
  Dirac fields string-localized, while the observable Maxwell fields
  remain point-localized. This remarkable structural resolution of the
  old locality issues of QED must have escaped the attention of the authors of
  \cite{BCRV} by their focus on the generation of charged states.

In the case of full QED with the string-localized interaction density
(\ref{Le}), we restrict ourselves to first-order calculations. The results
outlined before show that all the characteristic features of Gauss' Law in QED
can be obtained in this way. In addition, we shall present a remarkable
``semi-perturbative'' formula, Eq.\ (\ref{psiK}), that reveals how the
interacting Dirac field ``virtually'' (in a sense to be explained) carries the
longitudinal degrees of freedom of the Maxwell potential of other approaches.

The full construction of the interacting charged field and the study
of its large time asymptotic infra-particle scattering behaviour lie well beyond the scope of this
paper. 

\section{Bogoliubov's formula}

\label{s:BF} \setcounter{equation}{0} 

The perturbative approach is based on
Bogoliubov's formula  \cite[Chap.\ 17]{BS} via a causal S-matrix functional of
an IR cutoff function  $g(x)$ for the coupling constant. The formula assigns
to each free  or composite free field $\chi$ an interacting field
$\chi\big\vert_{gL}$ where $L$ is the interaction density. This map  deforms
the algebraic relations, but it respects local  commutativity of the
interacting observables, thanks of the  causal properties of the S-matrix
functional. The renormalization is  done in the Glaser-Epstein framework of
causal perturbation theory.

Bogoliubov's formula with an interaction density $L(x)$ reads 
\begin{align}
\label{bog}\int d^{4}x \, f(x)\chi\big\vert_{gL}(x) \equiv\chi\big\vert_{gL}(f) 
:= -i\partial_{\lambda}S[g,0]^{*}S[g,\lambda f]\Big\vert_{\lambda=0},
\end{align}
where the causal S-matrix is the time-ordered exponential
\[
S[g,f]:= T e^{i \int d^{4}x \, [g(x)L(x) + f(x)\chi(x)]}.
\]
In first order,
\begin{align}
\label{first}\chi\big\vert_{gL}(x)= \chi(x)+ \int d^{4}y \, g(y)R(L(y),\chi
(x)) + O(g^{2}),
\end{align}
where $-iR(A(y),B(x)):= T[B(x)A(y)]- A(y)B(x)$ is the retarded
commutator, and the higher order terms are multiple retarded
commutators \cite{DF}. By the Wick expansion, the causal S-matrix
$S[g,f]$ is
expanded into a sum of Wick products times numerical retarded
(anti-)commutators of free fields. Already these propagators are ill-defined
by the prescription ``$\theta(x^{0}-y^{0})[\phi(x),\chi(y)]_\pm$''
(being products of distributions with overlapping singular supports), even more so are products thereof appearing
in loop diagrams ill-defined. Their definition is a matter of
renormalization and may introduce free parameters to be fixed by
renormalization conditions. In Glaser-Epstein causal perturbation
theory, renormalization is done in position space so as to keep control of
localizations.

In the case of our interest, the interaction density (\ref{Le}) is
string-localized, where $L(e)$ must be averaged with a test function $h(e)$.
The renormalization is then subject to the ``principle of
string-independence'': the S-matrix and observable fields must be independent
of the auxiliary string variable $e$ or its averaging function $h$ in the
``adiabatic limit'' $g(x)\to q$.

The preservation of local commutativity of the observables is a crucial
feature of Bogoliubov's formula. Let us sketch (in first order) how the
argument proceeds, and why it fails for the non-observable Dirac fields in QED.

The commutator of two interacting fields $A_{1}$, $A_{2}$ in first order is
\[
\int d^{4}y \, g(y)C(y), \quad\hbox{where}\quad 
C(y)=[R(L(y),A_{1}),A_{2}]+[A_{1},R(L(y),A_{2})].
\]
For sets $X$, $Y\subset\RR^{4}$, we say that $X$ is later than $Y$
($X\succcurlyeq Y$) if for each $x\in X$ and $y\in Y$, $x$ is not in the
closure of the causal past of $y$. Two sets are spacelike separated iff
$X\succcurlyeq Y$ and $Y\succcurlyeq X$. Let $X_{i}$ and $Y$ be the
localizations of $A_{i}$ and $L(y)$, respectively. They may be points or strings.

Let the localizations $X_{i}$ of the free field operators $A_{i}$ be spacelike
separated. If $Y\succcurlyeq X_{1}$ and $Y\succcurlyeq X_{2}$, then
$R(L(y),A_{1})=0=R(L(y),A_{2})$ and hence $C(y)=0$. If $Y\not \later X_{1}$
and $Y\not \later X_{2}$, then the retarded commutators coincide with the
ordinary commutators, and $C(y)=0$ by the Jacobi identity. For the other two
cases, assume first that $Y=\{y\}$ is a point. If $y\succcurlyeq X_{1}$ and
$y\not \later X_{2}$, then $C(y) = i[A_{1},[L(y),A_{2}]]$. Now $y\not \later
X_{2}$, together with the fact that $X_{1}\succcurlyeq X_{2}$ implies by
\cite[Lemma 2.5]{CMV} that $X_{1}\succcurlyeq y$, and because also
$y\succcurlyeq X_{1}$, $y$ is spacelike separated from $X_{1}$, and $C(y)=0$
because $A_{1}$ commutes with $L(y)$ and with $A_{2}$. Similar for
$y\succcurlyeq X_{2}$ and $y\not \later X_{1}$. Hence $C(y)=0$ for all $y$.
Thus, a point-local interaction density preserves local commutativity of
string-localized fields.

Lemma 2.5 in \cite{CMV} does not apply when $Y$ is a string  $y+\RR_{+}e$,
even if $X_{i}$ are points. But when $A_{i}(x)$ are  observable fields, which,
by definition, remain independent of $e$ under the interaction, one can
exploit the option to choose $e$ appropriately, so that the same conclusion
holds: commutativity of point-local {\em observables} is preserved by string-local interactions. We shall present an alternative, more elegant argument in
 \sref{s:Hy}.

For non-observable fields, one has to expect an uncontrolled delocalization
under a string-localized interaction, in general. However, for the interacting
Dirac field in QED, we shall present a remarkable formula in \sref{s:Hy}
by which it behaves like a string-localized field under a point-localized
interaction. Thus, it ``inherits'' the string-localization of the interaction
density. This is in accord with the NoGo result of \cite{FPS} mentioned in the
introduction. If $e$ is chosen spacelike, this also ensures sufficient
spacelike separability for the needs of scattering theory.

\section{An external field warm-up}

\label{s:Ex} \setcounter{equation}{0} 

We present a simple example, where
perturbation theory is exact. The example is the external field problem with
${\mathcal{L}}_{h}=F^{\mu\nu}h_{\mu\nu}$ with an arbitrary classical source
$h_{\mu\nu}(x)$. We shall show that the Bogoliubov map changes the values of
the central flux operators. This shows that, even when the interaction density
is a functional of the observables, the Bogoliubov map is not an approximation
by inner conjugations.

In the case at hand, Bogoliubov's formula (\ref{bog}) simplifies to
\begin{align}
\label{Bog}\int d^{4}x \, F_{h}^{\mu\nu}(x)f_{\mu\nu}(x) \equiv F_{h}(f) =
-i\partial_{\lambda}S[h]^{*}S[h+\lambda f]\Big\vert_{\lambda=0},
\end{align}
where
\[
S[h]:= T e^{i \int d^{4}x \, {\mathcal{L}}_{h}(x)} = T e^{i F(h)}.
\]
(As the external field $h$ plays the role of the cutoff function $g$, an
adiabatic limit is not taken.)

The retarded propagator 
\[R(F(f),F(h)) = G_{\mathrm{ret}}(f,h)\cdot\eins \equiv \int d^{4}x \, d^{4}y\, f_{\mu\nu}(x)\,h_{\kappa
\lambda}(y) \,G^{\mu\nu,\kappa\lambda}_{\mathrm{ret}}(x-y)\cdot\eins
\]
is given in terms of the massless scalar propagator $G_{\mathrm{ret}0}(z)=-\int\frac{(2\pi)^{-4}d^{4}k}{k^{2}+i\eps k^{0}}e^{-ikz}$:
$$G_{\mathrm{ret}}^{\mu\nu,\kappa\lambda}(z) = \partial^{[\mu}\eta^{\nu][\lambda}\partial^{\kappa]} G_{\mathrm{ret}0}(z)$$
(the brackets denoting anti-symmetrization).
Thus, standard use of Wick's theorem implies
\[
S[h] = e^{\frac i2 G_{\mathrm{ret}}(h,h)}\cdot e^{i F(h)}
\]
and
\[
S[h]^{*}S[h+f] = e^{i G_{\mathrm{ret}}(f,h)+\frac i2 G_{\mathrm{ret}}(f,f)}\cdot e^{i F(f)}.
\]

Bogoliubov's formula then yields
\[
F_{h}(f)= F(f) + G_{\mathrm{ret}}(f,h)\cdot\eins, \quad\hbox{or}\quad
F_{h}^{\mu\nu}(x) = F^{\mu\nu}(x)+\int d^{4}y\, G_{\mathrm{ret}}^{\mu
\nu,\kappa\lambda}(x-y)h_{\kappa\lambda}(y)\cdot\eins.
\]

The string-localized interaction density $A^{\mu}(x,e)j_{\mu}(x)$ with a
classical conserved source $j_{\mu}(x)$ is a special case, because the action
can be written as
\begin{align}
\label{action}\int d^{4}x \, A^{\mu}(x,e)j_{\mu}(x) = F(h^{e}),
\end{align}
where $h^{e}_{\mu\nu}(x) = \frac12 \int_{0}^{\infty} ds\, j_{[\mu}(x-se)e_{\nu]}$.

When the Maxwell Green function is expressed in terms of the scalar Green
function, the shift term becomes
\begin{align}
\label{Gh}(G_{\mathrm{ret}}h^{e})^{\mu\nu}(x) \equiv \int d^{4}y\,
  G_{\mathrm{ret}0}(x-y)\,
  \eta^{\kappa[\mu}\partial^{\nu]}\partial^{\lambda}\int_{0}^{\infty} 
ds \, j_{[\kappa}(y-se)e_{\lambda]}.
\end{align}
Because $\partial^{\lambda}j_{\lambda}=0$ and $e_{\lambda}\partial^{\lambda
}j(y-se)=-\partial_{s} j(y-se)$, the derivative $\partial^{\lambda}$ of the
$s$-integral just yields $j_{\kappa}(y)$, which is independent of $e$.
Actually, the retarded propagator of the field strength is unique only up to a
term $c\cdot(\eta^{\mu\kappa}\eta^{\nu\lambda}-\eta^{\nu\kappa}\eta
^{\mu\lambda})\delta(x-x^{\prime})$. This term would add a contribution to
(\ref{Gh}) that depends on $e$. The principle of $e$-independence fixes the
renormalization constant $c=0$.

Thus, the shift term is independent of $e$ and equals the classical
retarded electromagnetic field with source $j^{\mu}$,
\begin{align}
\label{Fclass}F^{\mu\nu}_{\mathrm{class}}(x)=\int d^{4}y\, G_{\mathrm{ret}0}(x-y)\, \partial^{[\nu} j^{\mu]}(y).
\end{align}
In particular, the flux operator is shifted by the classical value of the flux.

No unobservable quantum degrees of freedom are needed to achieve this
result. This instance is in contrast to the statement in the abstract
of \cite{BCRV}, that ``[gauge] bridges are needed in order to ensure
the validity of Gauss' law''.

\section{Gauss' Law in string-localized QED}

\label{s:QED} \setcounter{equation}{0} 

The following explicit first-order
calculation of the interacting flux operator in string-localized QED shows
that Bogoliubov's formula changes the free flux operators to interacting flux
operators that satisfy Gauss' Law as an operator equation, with the electric
charge operator ``on the right-hand side'' plus a boundary term that weakly
vanishes in the adiabatic limit.

The string-localized interaction density is $L(x,e)=A_{\mu}(x,e)j^{\mu}(x)$,
where $j^{\mu}={:\overline\psi\gamma^{\mu}\psi:}$ is the Dirac current. We
choose for simplicity $e=(0,\vec e)$.

The interacting field strength in first order is
\[
F_{\mu\nu}\big\vert_{gL(e)}(x) = F_{\mu\nu}(x) + \int d^{4}y\,
g(y)\,R(A_{\kappa}(y),F_{\mu\nu}(x)) j^{\kappa}(y) + O(g^{2}).
\]
The retarded commutator arises from $R(F_{\kappa\lambda}(y+se),F_{\mu\nu
}(x))e^{\lambda}$ by distributional integration over the string according to
(\ref{slok}). $R(F,F)$ is unique up to a renormalization as in
\sref{s:Ex}. The choice $c=0$ yields
\[
R^{(0)}(A_{\kappa}(y),F_{\mu\nu}(x)) = -\Big(\partial^{x}_{[\mu}\eta
_{\nu]\kappa} + \partial^{x}_{[\mu}e_{\nu]} \partial_{\kappa}^{y} I^{y}_{e}\Big)
G_{\mathrm{ret}0}(x-y)\cdot \eins,
\]
where $(I_{e}^{y}f)(y):= \int_{0}^{\infty}ds\, f(y+se)$ is the string integration
operator with inverse $-(e\partial_{y})$. 

Its divergence is
\begin{align}
\label{divR}\partial_{x}^{\mu}R^{(0)}(A_{\kappa}(y),F_{\mu\nu}(x)) = \Big[
-\eta_{\nu\kappa}\delta(x-y)+e_{\nu}\partial^{x}_{\kappa}\int_{0}^{\infty}ds\,
\delta(x-y-se)\Big]\cdot\eins.
\end{align}
Thus,
\begin{align}
\label{gauss}\partial^{\mu}F_{\mu\nu}\big\vert_{gL(e)}(x) = - g(x) j_{\nu}(x)
+ e_{\nu}\int_{0}^{\infty}ds\, \partial_{\kappa}g(x-se)\cdot j^{\kappa}(x-se)
+ O(g^{2}).
\end{align}
The first term becomes $-q j_{\nu}(x)$ in the adiabatic limit. For spacelike
$e$, the second term, evaluated in an electron state $\Psi=\psi^{*}(f)\Omega$,
${\vert\!\vert\Psi\vert\!\vert}^{2}=1$, goes to zero in the adiabatic limit,
because as the region where $g(x)=q$ increases, the support of $\partial
_{\kappa}g(x)$ moves to infinity in a spacelike direction, where $(\Psi,
j^{\kappa}(x)\Psi)$ decays rapidly. Thus, the adiabatic limit exists in the
weak sense, and the differential Gauss Law holds weakly. In the adiabatic
limit,
\begin{align}
\label{glob}\int_{V} d^{3}x \,(\Psi, \vec\nabla\vec E\big\vert_{qL(e)}(0,\vec
x)\Psi) = - q \int_{V} d^{3}x\, (\Psi, j_{0}(0,\vec x)\Psi) + O(q^{2}).
\end{align}
If the volume $V$ is large enough that its complement is spacelike separated
from the support of $f$, this equals $-q$. In particular, also the global
Gauss Law (\ref{gGL}) holds.

Because $\partial_{\mu}F^{\mu\nu}=0$ in 0th order, the 1st order corrections
of $\psi(f)\big\vert_{qL(e)}$ do not contribute, and the previous are the full
first order results. Higher perturbative orders are needed to turn the current
on the right-hand side into the interacting current.

If the same calculations leading to (\ref{gauss}) were done in the point-local
indefinite-metric (Krein space) setting, for simplicity in the Feynman
gauge $\lambda=1$,
the zeroth-order term $\partial^{\mu}F_{\mu\nu}=-\partial_{\nu}(\partial A)$
would not vanish (the ``fictitious current'' mentioned in 
\sref{s:GL3}, cf.\ \aref{a:IM})
and must be added to (\ref{gauss}). We shall see in \sref{s:IP} that the
fictitious current contributes to the expectation value of $\partial_{\mu
}F^{\mu\nu}$ in states generated from the vacuum by the interacting Dirac
field, so as to cancel the global charge. In view of \cite{FPS}, this is
a necessity because the Dirac field is point-localized in indefinite-metric Feynman
gauge QED, and consequently commutes with the gobal charge
  operator. Conversely, in the string-localized approach, the global
charge is the expected one, and the charged fields are string-localized.

The bulk term $-g(x)j(x)$ in (\ref{gauss}) would be the same in the
indefinite-metric setting, but the boundary term would be instead
\begin{align}
\label{fict}\int d^{4}y\, g(y) j^{\kappa}(y) \,\partial_{\kappa}\partial_{\nu
}G_{\mathrm{ret}0}(x-y) = \int d^{4}y\, \partial_{\kappa}g(y) \cdot j^{\kappa}(y)
\partial_{\nu}G_{\mathrm{ret}0}(x-y).
\end{align}
This vanishes for large spacelike $x$ where the support of $G_{\mathrm{ret}0}(x-y)$ does
not intersect the support of $\partial g(y)$. If the integral of the
zero-component over the $x^{0}=0$-plane is computed before the adiabatic limit
is taken, the integral over $\partial_{0}G_{\mathrm{ret}0}(x-y)$ yields $\theta
(x^{0}-y^{0})$, and by partial integration the boundary term (\ref{fict})
exactly cancels the bulk term. On the other hand, the decay of (\ref{fict}) in
the adiabatic limit for finite $x$ is harder to control because the
integration extends over the intersection of the support of $\partial g$ with
the entire backward lightcone of $x$: the concentration of the boundary term
along the string in (\ref{gauss}) is a technical advantage of the
string-localized approach.

The propagator of the field strength, and consequently also the retarded
commutator, has a renormalization freedom
\[
R^{(c)}[F_{\mu\nu}(x),F_{\kappa\lambda}(y)]=R^{(0)}[F_{\mu\nu}(x),F_{\kappa
\lambda}(y)]+c\cdot(\eta_{\mu\kappa}\eta_{\nu\lambda}-\eta_{\nu\kappa}
\eta_{\mu\lambda})\delta(x-y)\cdot\eins.
\]
Integration over the string $e=(0,\vec e)$ gives the corresponding freedom for
the retarded commutator $R(A_{\kappa}(y),F_{\mu\nu}(x))$ of the electric field
with the string-localized potential
\[
-ic\cdot\eta_{\kappa[\mu} e_{\nu]}\int_{0}^{\infty}ds\,\delta(x-y-se).
\]
Its divergence is $c$ times (\ref{divR}). Thus, the renormalization freedom
just renormalizes the electric charge. In other words, if $q$ is the physical
unit of charge, then one must choose $c=0$, cf.\ also \cite{MS2}. The same
choice is also dictated by the principle of string-independence
(as in \sref{s:BF}).

\section{Infra-particles}

\label{s:IP} \setcounter{equation}{0}

In order that the global charge operator = electric flux through the
infinite sphere exists and is non-zero in charged states $\Psi$, the  field
strength should decay like $r^{-2}$ in spacelike directions
(corresponding to the classical Coulomb Law). Thus, for 
the interacting field the limit
\bea{fmunu}
f_{\mu\nu}(x):=\lim_{\lambda\to\infty}\lambda^{2} F_{\mu\nu}(\lambda
x)
\eea
for spacelike $x$ should exist in the weak sense (matrix elements),
be non-zero and have finite 
fluctuations in charged states \cite{Bu2}. Because $f_{\mu\nu}(x)$
commutes with all local observables, it is a multiple of $\eins$ in
every irreducible representation: $f_{\mu\nu}(x)=a_{\mu\nu}(x)\cdot\eins$. This 
asymptotic field configuration $a_{\mu\nu}(x)$ is by construction a homogeneous
function of $x$ of degree $-2$. Buchholz 
\cite{Bu2} has shown that these properties imply that 
charged states cannot be eigenvectors of the mass operator $M^2=P_\kappa
P^\kappa$. (For a simplified argument, see \aref{a:CO}.)

We conclude that the electrons of QED are ``infra-particles''
  (particles without a sharp mass), because Gauss' Law
  enforces the decay like $r^{-2}$ of the surrounding electromagnetic
  field (the ``photon cloud''). In the two-point function of fields
  describing infra-particles, the $\delta$
function $\delta (p^2-m^2)$ is replaced by some continuous function with
a singularity (cf.\ the 2D model in \cite{S63}). This in turn entails
that correlations decay faster in asymptotic time 
than with a sharp mass. For this reason, the scattering
theory for infra-particles needs a nontrivial adjustment of the LSZ formalism.

We want to verify the infra-particle features of QED in the
  string-localized approach. The validity of the global Gauss Law established in \sref{s:QED}
already entails the $\lambda^{-2}$ decay of the radial electric field (flux
density). We want to calculate also its directional distribution.

To this end, we have to evaluate the
asymptotic field configuration of the interacting field in charged states $\Psi\big\vert_{gL(e)}:=\psi^{*}\big\vert_{gL(e)}(f)\Omega$
created by the interacting Dirac field:
\[
a_{\mu\nu}(x)=\lim_{\lambda\to\infty}\lambda^{2} (\Psi,F_{\mu\nu}(\lambda
x)\Psi)\big\vert_{gL(e)}
\]
In contrast to indefinite-metric
approaches, this state defines a positive functional.

The three first-order contributions to $\psi$, $\overline\psi$ and $F$ in
$\big\langle \psi(f)F^{\mu\nu}(x)\psi(f)^{*}\big\rangle$ are in turn
\begin{align}
\label{X}X_{1}\equiv\big\langle\psi^{(1)}(f)F_{\mu\nu}(x)\psi(f)^{*}\big\rangle & =\int d^{4}y\,
g(y)\,\big\langle R(j^{\kappa}(y),\psi(f))\psi(f)^{*}\big\rangle\cdot
D_{\kappa,\mu\nu}^{y}\Delta_{0}(y-x),\nonumber\\
X_{2}\equiv\big\langle\psi(f)F_{\mu\nu}(x)\psi^{(1)}(f)^{*}\big\rangle & =\int d^{4}y\,
g(y)\,\big\langle \psi(f)R(j^{\kappa}(y),\psi(f)^{*})\big\rangle\cdot
D_{\kappa,\mu\nu}^{y}\Delta_{0}(x-y),\nonumber\\
X_{3}\equiv\big\langle\psi(f)F_{\mu\nu}^{(1)}(x)\psi(f)^{*}\big\rangle & =\int d^{4}y\,
g(y)\,\big\langle \psi(f)j^{\kappa}(y)\psi(f)^{*}\big\rangle \cdot
D_{\kappa,\mu\nu}^{y} G_{\mathrm{ret}0}(x-y),\nonumber
\end{align}
where $\Delta_{0}$ and $G_{\mathrm{ret}0}$ are the massless scalar 2-point functions and
retarded propagator. The tensor of integro-differential operators
\[D_{\kappa,\mu\nu}^{y}= (\partial_{\mu}^{y}\eta_{\nu\kappa}-\partial_{\nu}^{y}
\eta_{\mu\kappa}) + \partial_{\kappa}^{y}\,(\partial^{y}_{\mu}e_{\nu
}-\partial^{y}_{\nu}e_{\mu}) \, I^{y}_{e}
\]
arises from the two-point functions involving $F_{\mu\nu}(x)$ and $A_{\kappa
}(y,e)=I^{y}_{e}e^{\lambda}F_{\kappa\lambda}(y)$. Notice the split-up into a
string-independent part (the only one that would be present in the
indefinite-metric approach in the Feynman gauge) and a string-dependent part.

With a hindsight from \sref{s:Hy} and \ref{s:Phot}, we anticipate that
the relevant contributions in the asymptotic limit arise only from the
string-dependent parts of $D_{\kappa,\mu\nu}$ in $X_{1}$ and $X_{2}$, while
the sum of all other contributions decays faster than $\lambda^{-2}$. We defer
the proof of the latter statement to the end of the section.

In the relevant contributions, we partially integrate $\partial^{y}_{\kappa}$,
using
\begin{align}
\label{dRj}\partial^{y}_{\kappa}\,R(j^{\kappa}(y),\psi(f))=-if(y)\cdot\psi(y),\quad
\partial^{y}_{\kappa}\,R(j^{\kappa}(y),\psi(f)^{*})=i\overline{f(y)}\cdot
  \psi^{*}(y).
\end{align}
The boundary terms involving $\partial_{\kappa}g(y)$ can be roughly estimated
to vanish  in the adiabatic limit like $T^{-3}$ if the cutoff function $g(y)$
is chosen to drop from $q$ to $0$ in the interval $T\leq\vert y^{0}\vert\leq
T+1$. Specifically, the support of the retarded  commutators is the backward
lightcone of the support of  $f$, and intersects the support of $\partial g$
in a strip of  spatial volume $\sim T^{3}$. Taking into account the quadratic
decay  of two-point functions and propagators in timelike directions, the
boundary terms are $\sim T^{3}/T^{6}$.

The bulk terms are
\[
i\int d^{4}y\,g(y)\Big(f(y)\cdot(\partial_{y}\wedge e)I^{y}_{e}\Delta
_{0}(y-x)\cdot\big\langle \psi(y)\psi(f)^{*}\big\rangle - \big\langle \psi
(f)\psi^{*}(y)\big\rangle\cdot(\partial_{y}\wedge e) I^{y}_{e}\Delta
_{0}(x-y)\cdot\overline{f(y)} \Big).
\]
In the adiabatic limit, we may assume that the support of $f$ is contained in
the region where $g(y)=q$, and by the scaling behaviour $\lambda^{2}\Delta
_{0}(\lambda x-y)=\Delta_{0}(x-y/\lambda)$ of the massless two-point function,
we may simply neglect $y\in\mathrm{supp}(f)$ against $x$. We obtain in first
order
\begin{align}
\label{IP}a(x)^{(1)}=\lim_{\lambda\to\infty} \lambda^{2} \big\langle \psi(f)
F(\lambda x) \psi(f)^{*}\big\rangle\big\vert_{gL(e)}^{(1)} = q\cdot
(\partial_{x}\wedge e) \, I^{x}_{-e} C_{0}(x)\cdot\big\langle \psi
(f)\psi(f)^{*}\big\rangle,
\end{align}
where $C_{0}$ is the massless commutator function. The last factor is
${\vert\!\vert\Psi\vert\!\vert}^{2}=1$. If we choose $e^{0}=0$, then $a(0,\vec
x)^{(1)}$ evaluated in the plane $x^{0}=0$ vanishes for the magnetic field,
and
\begin{align}
\label{elec}\lim_{\lambda\to\infty} \lambda^{2} \big\langle \psi(f) \vec
E(\lambda\vec x) \psi(f)^{*}\big\rangle = -q\cdot\vec e \int_{0}^{\infty
}ds\,\delta(\vec x-s\vec e).
\end{align}
for the electric field. Thus the asymptotic electric field is concentrated at
the direction $\vec e$ and points in the direction of $-\vec e$. The total
flux is $-q$, as expected for the electron state $\Psi$, by the global Gauss
Law, and in accord with \sref{s:QED}. If $L(e)$ is averaged with a
smearing function $h(e)$, the asymptotic flux density in the direction $\vec
x=r\vec e$ is $-qh(e)/r^{2}$.

To show the vanishing of the sum of the remaining contributions $X_{3}$ and
the string-independent parts of $X_{1}$, $X_{2}$, we content ourselves with
showing that their contribution to  $\partial^{\mu}\big\langle \psi
(f)F_{\mu\nu}\psi(f)^{*}\big\rangle$ is a sufficiently  well-localized charge
distribution of total charge zero, so its asymptotic field configuration
vanishes. We use
\begin{align}
\label{nn}\partial^{x\mu} (\partial_{\mu}^{y}\eta_{\nu\kappa}-
\partial_{\nu}^{y}\eta_{\mu\kappa}) \Delta_{0} & = (-\eta_{\nu\kappa}\square^{y} 
+\partial^{y}_{\nu}\partial^{y}_{\kappa})\Delta_{0}
=\partial^{y}_{\nu} \partial^{y}_{\kappa}\,\Delta_{0},\nonumber\\[1.3mm]
\partial^{x\mu} D^{y}_{\kappa,\mu\nu}G_{\mathrm{ret}0} & = -\big(\eta_{\nu\kappa} +
e_{\nu}I^{y}_{e}\partial^{y}_{\kappa}\big)\square^{y}G_{\mathrm{ret}0}= -\big(\eta
_{\nu\kappa} + e_{\nu}I^{y}_{e}\partial^{y}_{\kappa}\big)\delta(x-y),\nonumber
\end{align}
in the string-independent parts of $X_{1}$ and $X_{2}$, and in $X_{3}$, respectively.

Partial integration of $\partial^{y}_{\kappa}$ in the former produces bulk
terms from the action on the propagators, as before, and boundary terms that
identically cancel the corresponding contributions of the boundary terms in
the string-dependent parts (because $-(e\partial)$ inverts the
string-integration $I_{e}$). Partial integration of $\partial^{y}_{\kappa}$ in
the latter produces only a boundary term that is identical with the rapidly
decaying boundary term in (\ref{gauss}).

Having settled the boundary terms, we collect the bulk terms:
\[
q\cdot\int dx_{1}dx_{2} f(x_{1})\overline{f(x_{2})}\partial^{x}_{\nu
}\Big(-i\Delta_{0}(x_{1}-x) +i\Delta_{0}(x-x_{2})\Big)\cdot\big\langle \psi
(x_{1})\psi^{*}(x_{2})\big\rangle 
- g(x) \big\langle \psi(f) j_{\nu}(x)\psi(f)^{*}\big\rangle.
\]
For large $x$, one may again neglect $x_{i}$ in the arguments of $\Delta_{0}$.
For $\nu=0$ and $x^{0}=0$, the first two terms combine into $\partial^{0}_{x}
C_{0}(x)=\delta(\vec x)$. The resulting contribution to the charge density in
the asymptotic limit is
\begin{align}
\label{comp}\big\langle \psi(f) \big[q\delta(\vec x)-g(x)j^{0}(0,\vec
x)\big]\psi(f)^{*}\big\rangle.
\end{align}
This result exhibits a compensating point charge $q\delta(\vec x)$ that one
  recognizes as coming from the ``fictitious current''. Indeed, its ``position at $x=0$'' is
fictitious because it is ``seen from infinity''. In the Feynman gauge
calculation, the string-dependent part \eref{elec} would be absent
and the total result would be  (\ref{comp}).

The total charge in (\ref{comp}) vanishes in the adiabatic limit, hence the
contribution to the asymptotic field configuration vanishes, too. Multi-pole
radiation fields (that are not excluded by the vanishing total charge, and
have a slower spatial decay) also have zero asymptotic field  configurations,
because the precise definition of the asymptotic  limit \cite{Bu2} involves
also an averaging in time that suppresses  the oscillations.

\section{The hybrid approach}

\label{s:Hy} \setcounter{equation}{0} 

The ``hybrid approach'' \cite{MS} allows
to  study the relation between the string-localized and the  indefinite-metric
approach. In particular, it sheds light  on how the superselection structure
of QED arises dynamically (\sref{s:Phot}).

Let $A_\mu(x)$ be the usual point-localized vector potential, for
simplicity in the Feynman gauge. In order to emphasize that it is defined only in a 
space with indefinite metric (Krein space, see \aref{a:IM}), we denote it as $A^{K}_{\mu}(x)$. It can be decomposed as
\begin{align}
\label{AKA}A^{K}_{\mu}(x)=A_{\mu}(x,e)-\partial_{\mu}\phi^{K}(x,e).
\end{align}
The string-localized potential $A_{\mu}(x,e)$, defined as the string integral
(\ref{slok}) over the field strength, directly descends to the physical
Hilbert space, whereas the massless ``escort field''
\begin{align}
\label{phi}\phi^{K}(x,e):= \int_{0}^{\infty}ds\, A^{K}_{\mu}(x+se)e^{\mu}
\end{align}
lives on the Krein space. (The two parts of the operator $D_{\kappa,\mu\nu}$
in \sref{s:IP} precisely correspond to $A^{K}_{\mu}(x)$ and
$\partial_{\mu}\phi^{K}(x,e)$, respectively.)

The interaction density splits accordingly as
\begin{align}
\label{LKL}L^{K}=A^{K}_{\mu}j^{\mu}= A_{\mu}(e)j^{\mu}- \partial_{\mu
}\big[\phi^{K}(e)j^{\mu}\big].
\end{align}
$L(e)=A_{\mu}(e)j^{\mu}$ thus differs from $L^{K}$ by a total derivative that
should be ineffective in the adiabatic limit. $L(e)$ is a priori defined on
the Krein space, but descends to the physical Hilbert space, while the
indefinite-metric degrees of freedom are ``disposed of'' with the discarded
total derivative. We have checked up to second order \cite{MS} that the $S$
matrix with interaction density $gL^{K}-\partial_{\mu}g\cdot\phi^{K}(e)j^{\mu
}$ coincides with the $S$ matrix with interaction $g L(e)$, i.e., the former
descends to the Hilbert space where the latter is defined.

This pattern prevails in many models of interest: there is a
string-independent point-localized interaction density $L^{p}$, possibly on a
Krein space, such that $L(e)$ descends to the physical Hilbert space and
$L^{p}-L(e)$ is a total derivative:
\[
L(e)= L^{p} + \partial_{\mu}V^{\mu}(e).
\]
By definition, an interacting field is observable if and only if
\begin{align}
\label{obs}\chi\big\vert_{gL(e)}=\chi\big\vert_{gL^{p}-\partial g\cdot V(e)}.
\end{align}
In particular, the left-hand side is defined on the Hilbert space and the
right-hand side, in the adiabatic limit $\partial g=0$, does not depend on the
string $e$ of the interaction density and is local because $L^{p}$ is
point-localized, by the argument given in \sref{s:BF}. By equality,
$\chi\big\vert_{gL(e)}$ enjoys both properties. This broad definition also
includes cases like $A(x,e^{\prime})\big\vert_{L(e)}$ in QED, that satisfies
(\ref{obs}) and hence is independent of $e$ and remains localized along the
string $e^{\prime}$.

For the interacting Dirac field in QED, a remarkable formula is expected to hold: 
\begin{align}
\label{psiK}\psi(x)\big\vert_{gL(e)} = {:}e^{ig\phi^{K}(x,e)}\psi
(x){:}\big\vert_{gL^{K}-\partial g \cdot j\phi^K}.
\end{align}
(We have explicitly verified it up to second order \cite{MS}. In first order, it is equivalent to 
$$\int d^{4}y\,R(j^{\mu}(y),\psi
(x))\Big[g(y)\big(A_{\mu}(y,e)-A_{\mu}^{K}(y)\big) + \partial_\mu g(y)\phi^K(y)\Big] = ig(x)\phi
^{K}(x,e)\psi(x),$$ which is true by (\ref{AKA}) upon partial integration,
using (\ref{dRj}).) The same holds if both $L(e)$ and $\phi^{K}(x,e)$ are
smeared with a test function $h(e)$.\footnote{Since the free operator
  on the right-hand side of (\ref{psiK})
is a special case of \cite[Eq.\ (2.6)]{M} and \cite[Eq.\ (1.1)]{St}, the
string-localized interaction on the left-hand side naturally implements these
previous ideas of ``QED in terms of gauge invariant fields''.} Let us
discuss its consequences.

The left-hand side is defined on the Hilbert space. Thus, it allows to
define positive states on the algebra of the
Maxwell field of the form $\omega(X)=(\Omega,
\psi\vert_{gL(e)} X\psi\vert_{gL(e)}^{*}\Omega)$ before the adiabatic 
limit $g(y)\to q$ is taken, and these states remain positive in the
adiabatic limit. Thus, the interacting Dirac field complies with
Hilbert space positivity.

On the other hand, $\psi\big\vert_{gL(e)}$ is a priori badly
delocalized in the adiabatic limit due to the string-localized interaction density (cf.\
\sref{s:BF}). But the right-hand side is manifestly
string-localized (because with $\partial g=0$,
$e$ appears only in the free field, and not in the interaction
density). By equality, the interacting Dirac field is
string-localized -- without being just a string-integral over a
point-localized field as (\ref{slok}).

String-localization is the best one may expect for the charged interacting
Dirac field (cf.\ \cite{BF} for theories with a mass gap), and it is
physically essential because string-localized fields do not commute with the
asymptotic flux operators that measure the total charge
\cite{FPS}. It also secures enough causal separability for the needs
of scattering theory.

Notice the ``semi-perturbative'' nature of (\ref{psiK}), where  the
exponential already involves a partial summation in the coupling constant. It is this feature  that
allows to discern the emergence of superselection sectors in the next section.

The appearance of the exponential of the escort field on the right-hand side
is also interesting in the context of the questions raised in \cite{BCRV}.
Ignoring for the moment their singular nature, we note that operators like
$\psi\big\vert_{qL(e)}(x_{1})\psi^{*}\big\vert_{qL(e)}(x_{2})$ involve, via
(\ref{psiK}),
\begin{align}
\label{phi12}e^{iq (\phi^{K}(x_{1},e)-\phi^{K}(x_{2},e))}.
\end{align}
If (for simplicity) the string $e$ is chosen parallel to a straight line
$\gamma$ from $x_{1}$ to $x_{2}$, it holds
\begin{align}
\label{IAphi}\int_{\gamma}d\vec x\cdot\vec A^{K}(\vec x) = \phi^{K}(\vec
x_{1},e)-\phi^{K}(\vec x_{2},e),
\end{align}
so that the operators (\ref{phi12}) coincide with the unitaries (``gauge
bridges'') used in \cite{BCRV} to implement the local Gauss Law. In this
guise, via the equality (\ref{psiK}), the longitudinal degrees of freedom are
``virtually present'' in the string-localized QED on the Hilbert space. The
next section puts these formal considerations onto a more solid ground.

\section{Photon cloud superselection}
\label{s:Phot} \setcounter{equation}{0} 

The exponentials of the free escort
field appearing in (\ref{psiK}) are highly singular objects. We shall
demonstrate how they can be regularized in such a way that (a) Hilbert space
positivity is guaranteed (despite their original definition on the Krein
space), and (b) they generate states with ``photon clouds'', and
are therefore responsible for the uncountable superselection
structure \cite{Bu1}. Thus, we believe them to be the ``carriers'' of
the infinitely many asymptotic symmetries of the ``infrared triangle''
\cite{Str}. (The multiplying Dirac field in  (\ref{psiK}) plays no
essential role in the argument and will 
be omitted from our simplified presentation.) 

The photon clouds are characterized by the expectation values of
asymptotic field operators as in \sref{s:IP}, that in turn define
uncountably many superselection sectors. The method below was first
used in a $1+1$-dimensional model in order to understand the
appearance of infra-particles \cite{S63}. Its application to QED is
  a new result. It shows along the way that the model was on the right
  track, even if its massless particles were not the photons of a
  gauge theory. The crucial fact is that the infrared 
  state vectors do not exist in the original photon Fock space
  (tensored with the Dirac Fock space), but in a GNS
  Hilbert space reconstructed from the state functional, as it was also
  emphasized in Steinmann's approach \cite{St2} to QED. 

The escort field itself is singular due to the logarithmic divergence of its
two-point function. We first regularize it by introducing a mass $m$:
\begin{align}
\label{wm}w_{m}(x-x^{\prime},e,e^{\prime}) = \int_{H_{m}^{+}} d\mu_{m}(p)\,
e^{-ip(x-x^{\prime})} \frac{-e\cdot e^{\prime}}{(p\cdot e-i\eps)(p\cdot
e^{\prime}+i\eps)},
\end{align}
where $d\mu_{m}(p)=(2\pi)^{-3}d^{4}p\,\delta(p^{2}-m^{2})\theta(p^{0})$. When
the string directions are smeared with real test functions on the hyperboloid
$e^{2}=-1$, one gets
\[
w_{m}(x-x^{\prime},h,h^{\prime}) = \int_{H_{m}^{+}} d\mu_{m}(p)\,
e^{-ip(x-x^{\prime})} [-\overline{t(p,h)_{\mu}}t(p,h^{\prime\mu}],
\]
where
\[
t(p,h)^{\mu}:=\int_{0}^{\infty}ds \int_{S^{2}} d\sigma(e)\,e^{ip\cdot se} \,
h(e)e^{\mu}= i\int_{S^{2}} d\sigma(e)\,\frac{h(e)e^{\mu}}{(p\cdot e)+i\eps}.
\]
The positivity of $-\overline{t(p,h)_{\mu}}t(p,h)^{\mu}$ can be guaranteed by
restricting the support of $h$ to the sphere $e=(0,\vec e)$ (or any Lorentz
transform of it), so that $t(p,h)=(0,\vec t(p,h))$, and $-\overline
{t(p,h)_{\mu}}t(p,h)^{\mu}=\vert\vec t(p,h)\vert^{2}\geq0$.

The massless limit of the distribution $w(z,h,h^{\prime})$ is defined only for
test functions $g(z)$ with $\int g(z)\,d^{4}z= \widehat g(0)= 0$. In order to
enable the massless limit for arbitrary test functions, we define
\begin{align}
\label{wreg}w_{m,\mathrm{reg}}(g,h,h^{\prime})  & := \int_{H_{m}^{+}} d\mu
_{m}(p)\, [\widehat g(p) - \widehat g(0)v(p)]\, \overline{\vec t(p,h)}\vec
t(p,h^{\prime})\nonumber\\
& = w_{m}(g,h,h^{\prime})-\widehat g(0)\cdot c_{m,v}(h,h^{\prime}),
\end{align}
where $v$ is any test function with $v(0)=1$. A different choice of $v$ leads
to an additive constant, and one can see that this is the only freedom of
renormalization. On test functions with $\widehat g(0)\neq0$, both
$w_{m}(g,h,h^{\prime})$ and $\widehat g(0)\cdot c_{m,v}(h,h^{\prime})$ diverge
in the massless limit, but their difference is finite. Due to the subtraction,
$w_{m,\mathrm{reg}}(x-x^{\prime},h,h)$ is no longer positive.

We define the regularized exponential $e^{iq\phi^{K}(x,e)}$ of the escort
field as
\[
V_{m}(f,h):= e^{-\frac12 q^{2}\,c_{m,v}(h,h)} \cdot{:}\exp i\phi^{K}_{m}(f,h){:},
\]
where the real test functions have total weights $\int d^{4}x\,f(x)=q$ and
$\int_{S^{2}} d\sigma(\vec e)h(e)=1$. $V_{m}$ is defined on the GNS Hilbert
space of the positive two-point function $w_{m}(x-x^{\prime},e,e^{\prime})$.
The massless limit can be taken as follows.

Let $g(z):=\int dy\, f(z+y)f(y)$, hence $\widehat g(0)=q^{2}$. By Wick's
theorem,
\begin{align}
\label{VV}\big\langle V_{m}(f,h)V_{m}(-f,h^{\prime})\big\rangle =
\exp\Big[w_{m}(g,h,h^{\prime})-\frac12 q^{2}\, c_{m,v}(h,h)-\frac12 q^{2}\,
c_{m,v}(h^{\prime},h^{\prime})\Big]=\quad\nonumber\\
=e^{i\alpha(g,h,h^{\prime})}
\exp\Big[-\frac12 w_{m}(g,h-h^{\prime},h-h^{\prime})\Big]
e^{\frac12w_{m,\mathrm{reg}}(g,h,h)+\frac12
w_{m,\mathrm{reg}}(g,h^{\prime},h^{\prime})},\quad
\end{align}
where $\alpha(g,h,h^{\prime}) = \mathrm{Im}\, w_{m}(g,h,h^{\prime})$.

Because $\widehat g(0)\neq0$, $w_{m}(g,h-h^{\prime},h-h^{\prime})\to+\infty$
diverges and consequently
\[
\big\langle V_{m}(f,h)V_{m}(-f,h^{\prime})\big\rangle\to0
\]
in the limit $m\to0$, unless for all $p$
\begin{align}
\label{t2}\vert\vec t(p,h-h^{\prime})\vert^2=0.
\end{align}
Write $\vec H(\vec e)=(h-h^{\prime})(0,\vec e)\vec e$. Then
\begin{align}
\label{Ht}\frac 1{(2\pi)^3}\int d^3p \,e^{i\vec p\cdot r\vec e} \,\vec t(p,H) = \int_{S^2} d\sigma(\vec
e\,') \int_0^\infty ds\,\delta(r\vec e-s\vec e\,')\, \vec H(\vec e\,')
=r^{-2}\cdot \vec H(\vec e),
\end{align}
and (\ref{t2}) is equivalent to $h^{\prime}=h$. Thus,
\begin{align}
\label{sup}\big\langle V(f,h)V(-f,h^{\prime})\big\rangle = \Big\{
\begin{array}[c]{ll}
0 & \hbox{if $h'\neq h$}\\
\lim_{m\to0} e^{w_{m,\mathrm{reg}}(g,h,h)} > 0 \quad & \hbox{if $h'= h$.}
\end{array}
\end{align}
As limits of states (after
  insertion of field operators), $\langle V(f,h)\dots V(-f,h)\rangle$ are 
states on the free Maxwell field algebra. Because of the
orthogonality \eref{sup}, the GNS construction  (cf.\ \aref{a:CO})
yields uncountably many superselection sectors labelled by the
directional smearing functions $h$, exhibiting the expected breakdown
of Lorentz invariance \cite{FMS}, cf.\ \sref{s:GL2}.

We now compute the expectation values of the electromagnetic field strength in
states of charge $-q$ implemented by the regularized exponentials $V(-f,h)$ of
the free escort field. Again, we omit the  multiplying Dirac field. By
reproducing the same asymptotic field configurations determined by the
directional smearing function $h$ as in \sref{s:IP}, we see that only
these exponential fields are responsible for the photon clouds.

In the sequel, $F$ and $\phi^{K}$ are free fields, hence their commutator is a
multiple of $\eins$. The state obtained from the vacuum by the adjoint action
of $V(f,h)$
\[
\big\langle F_{\mu\nu}(x)\big\rangle_{f,h} := \frac{\big\langle V(f,h)F_{\mu
\nu}(x)V(-f,h)\big\rangle}{\big\langle V(f,h)V(-f,h)\big\rangle}= -i[F_{\mu
\nu}(x),\phi^{K}(f,h)]
\]
differs from the vacuum by the automorphism $\beta(F)=F-i[F,\phi^{K}(f,h)]$. (The two terms of the commutator correspond to the asymptotic
limits of the  string-dependent parts of $X_{1}$ and $X_{2}$ in the
calculation  in \sref{s:IP}. The remaining terms that were  shown in
\sref{s:IP} not to contribute asymptotically anyway, are  absent here
because we ignore the multiplying Dirac field.) The commutator is $[F_{\mu\nu
}(x),\phi^{K}(y,e)]=i(\partial^{x}\wedge e)I_{e}^{y}  C_{0}(x-y))$ in accord
with (\ref{IP}), smeared with $f(y)$ and $h(e)$. Repeating the calculation
after (\ref{IP}), gives  the same result
\begin{align}
\label{flux}\lim_{r\to\infty} r^{2}\big\langle \vec E(re)\big\rangle_{f,h} =
-q \cdot h(e)\vec e .
\end{align}
In other words, operators $V(f,h)$ with test functions $f(x)$  and $h(e)$ as
specified, substitute the singular expression  $e^{iq\phi^{K}(x,h)}$, that is
only perturbatively defined in  (\ref{psiK}) (with $\psi\big\vert_{L(h)}$ on
the left-hand side). They yield charged states with the asymptotic electric
flux density $-q h(e)/r^{2}$ in accord with the global Gauss Law, and with the
first-order result (\ref{elec}) of full QED. Perturbing the Dirac field with
different averages $L(h)=\int d\sigma(e)\,h(e)L(e)$, one can construct states
with arbitrary ``photon clouds'' whose shape is given by the function $h$.

For a full treatment, the interacting Dirac field must be taken
  into account, and not only its ``factor'' $e^{iq\phi^K}$ in
  (\ref{psiK}). It is clear from the above indirect
  construction of the exponential in a GNS representation, that
  (\ref{psiK}) does not remain a tensor product in the interacting
  field: it is impossible to split ``infrared matter'' into charged
  particles and their photon clouds. This important algebraic message goes well
  beyond more formal treatments as in \cite{Str}.  
  With a hindsight from the 2D model \cite{S63}, it
  is expected that the presence of this ``factor'' will 
reduce the large-time fall-off below the kinematic LSZ fall-off. As a
consequence, and in accord with \cite{YFS}, cross sections with
zero photon resolution vanish, and one has to resort to
the Bloch-Nordsieck prescription (soft-photon inclusive cross
sections, \cite{BN}). We hope to return to
these issues in a separate paper.

Quite analogous results concerning the
localization of the asymptotic flux and the inequivalence of representations, derived in
an external field setting, have been reported in \cite{DW}.
Interestingly, while in our approach the choice of the
  string-dependent interacting Dirac field creating the charged state
  is responsible for the string-dependent expectation value of the
  string-independent field operators,  in \cite{DW} the string-dependence is
  attributed to the field operator itself, via the choice of an axial
gauge condition in Dirac's quantization prescription.

\section{Conclusions}

\label{s:Conc} \setcounter{equation}{0} Gauss' Law is of  eminent importance
in QED. Its impact on the  algebraic structure and the nature of charged
particles seems to be  the decisive feature that distinguishes (abelian) gauge
theories \cite{H}.

We have presented several explicit perturbative calculations, showing that the
construction of QED with the help of string-localized potentials does
implement Gauss' Law in states with local charge distributions. In particular,
the result in \cite{BCRV} that in QED (once it is constructed),
string-localized potentials cannot be used for the implementation of ``gauge
bridges'', is not an argument against the possibility of the perturbative
construction of QED. To the contrary, Bogoliubov's formula applied to the
string-localized interaction density implies Maxwell's equations and, by
turning local free charged fields into string-localized interacting fields,
resolves the well-known conflict \cite{FPS} that the global Gauss Law cannot hold in a
QED with local charged fields.

The mechanism rendering charged fields string-localized when
  coupled to string-localized potentials, while
  preserving the point-localization of observables, is perhaps the
  most interesting achievement of these potentials. It was
not anticipated when they were originally proposed \cite{MSY,S,S2}
in order to improve the UV behaviour of perturbation theory. Their
role is certainly not to provide the degrees of freedom necessary to construct
charged states -- which they cannot as emphasized by \cite{BCRV}.

The string-localized potentials themselves appear
only in the interaction density: The quantities of interest of the resulting
theory are the local interacting field strengths (along with their fluxes) and
the string-localized interacting matter fields (along with their
point-localized currents),
related by the interacting Maxwell equation (\ref{inhmax}).

We acknowledge that the paper \cite{BCRV} has made it clear
that the quantum implementation of Gauss' Law requires degrees of freedom
beyond the observables. We have presented a ``hybrid formulation'' that makes
explicit the relation between the indefinite-metric and string-localized
approaches. It reveals in particular how these degrees of freedom are
``virtually present'' (in the sense explained in \sref{s:Hy}) in the
guise of escort fields. Their role is to mediate between gauge theory and
string-localized quantum field theory, by supplementing gauge theoretic
observables with string-localized charge-carrying interpolating fields in a
positivity-preserving way.

We have also shown that the approach satisfies Buchholz' 
infra-particle criterium (quadratic decay of the electric flux
density) in first order, with the asymptotic  field
configuration specified by the direction of the  string. Interestingly, what
might seem to be just a gauge degree of  freedom, becomes a feature of the
charged state created  by the Dirac field. Moreover, string-localized QED has
an infrared  mechanism (the 4D version of a mechanism first studied by one of
us in a 2D model \cite{S63}) to understand the superselection structure of
QED  due to asymptotic photon clouds.

The perturbative construction of QED is not done on the  observables
separately. Instead, we  perturb the Maxwell and Dirac fields simultaneously.
The former, being observables, are then distinguished by remaining local
under the  string-localized perturbation, while the latter become genuinely
string-localized. In this way, the dynamics of the theory itself ``selects''
the  interacting observables (including the currents) in terms of  their
causal localization  properties. Concerning the currents, the hybrid picture
and (\ref{obs}) are the tool to establish the expectation that
they are point-localized observables. The exponentiated escort fields in
(\ref{psiK}) intrinsically provide the ``gauge bridges'' that Brandt \cite{Br}
constructed in terms of potentials with longitudinal degrees of freedom.

\section{Outlook}

We conclude with some remarks, tracing out the perspective that is expected to
emerge from string-localized QFT (SLF) in more general theories.

Approaching a theory from different directions may better reveal  its ``inner
makings''. This is particularly worthwhile for gauge theories that hitherto
seem to defy the framework of Local Quantum Physics \cite{H}.

SLF can be successfully applied to large classes of models beyond QED
\cite{S2}, including SM weak interactions \cite{GMV}. We consider it as a
promising alternative to the gauge theory plus BRST setting, that does not
explicitly use indefinite metric at intermediate steps. In many
  important instances, it produces superficially the same results as gauge
  theory, e.g., comparing \cite{GMV} with \cite{DS}, but remains more
  economic (no ghosts) and physically transparent (e.g., massive vector
  bosons are massive from the outset). Its ``hybrid'' 
description (\sref{s:Hy}) provides the laboratory necessary to recognize
the relation (and hopefully equivalence) to the corresponding theory in the
usual approach, and to explain the ``miracle'' why Gauge Theory despite its
violation of positivity is so incredibly successful. Specifically, it allows
to control the localization of the charged fields in QED. Through the
exponentiation of the logarithmically divergent ``escort field'', it promises
a new interpretation of the Bloch-Nordsieck prescriptions in scattering
theory, by shifting the emphasis from momentum space to causal localization in spacetime.

In theories with massive vector particles, SLF allows to  maintain
renormalizability of the interaction with the Dirac  current, because
string-localized massive vector fields have a  better short-distance behaviour
than the Proca field. On the  other hand, a version of the hybrid approach
allows to use SLF  to control positivity if massive QED is described by the
renormalizable interaction with the indefinite massive Feynman gauge  potential.

SLF extends gauge theory in the sense that gauge theoretical observables are
complemented with gauge-invariant interpolating fields (in fact, in the hybrid
approach this is literally what happens). They are subject to the same
spectral analyticity properties of local QFT, and the fundamental
Spin-Statistics and PCT Theorems also apply. This has the enormous benefit of
providing a natural construction of particle states and scattering amplitudes
for which the analytic on-shell properties are a consequence of 
causal separability of interpolating fields, as envisaged by the pioneers of
dispersion relations in particle physics.

SLF becomes essential, and goes beyond the usual Lagrangean
approaches, whenever 
particles of spin (or helicity) $\geq1$ are  involved. This fact is linked to
the issue of symmetries \cite{O} that is much less subtle when only $s=0$ or
$s=\frac12$ particles are present.

A prominent instance is ``charge screening'' in theories with
massive vector bosons: the expectation value of the global charge
operator associated with a local conserved current necessarily vanishes \cite{Sw}. (Notice that this fact is at variance with the idea of
``spontaneous symmetry breaking'' which would require a divergent
expectation value.)\footnote{This conflict with the terminology of the Higgs
mechanism does not enjoy the attention that it deserves. The absence
of SSB does not mean that there is no Higgs particle -- only it
is not the driving agent, but its presence is a necessity for the
self-consistency of the theory, both in gauge theory \cite{ADS} and
in SLF \cite{S2,MS}.} It is presently not clear whether the charge
screening effect extends to non-Abelian currents coupled to massive
vector fields in SLF, so as to explain why Nature does not provide examples with exact particle multiplets
(besides PCT). Specifically, as in gauge theory non-Abelian currents
will not be gauge-invarint, they may not be local observables in
SLF. This important question deserves further attention.

Thinking of SLF as an alternative to gauge theory, can only work if it is
equally successful in determining the  ``correct'' SM interactions. In
contrast to interactions between particles of spin $0$ or $\frac12$, where the
model-building physicist can choose interactions ``at will'', interactions
involving particles of spin or helicity $>1$ are strongly constrained by
causality and positivity. Indeed, in SLF with several massive or massless
spin-one particles, like the electro-weak interactions, the Lie algebra
structure arises not by a symmetry principle, but instead (via the principle
of string-independence) as a consistency condition on the interaction density.

We expect that the new understanding of QED in terms of SLF will
also shed new light on helicity 2: perturbative gravity. 
Specifically, SLF will provide a new look at the asymptotic
(Bondi-Metzner-Sachs) symmetries at $h=2$ that avoids the NoGo
theorems in point-local gauge theory.

SLF may eventually help to extend the conceptional framework of  Axiomatic
QFT, without giving away its roots in  Locality and Positivity \cite{H}.
Especially Algebraic QFT is designed with the focus only on the observables,
and regards charged fields as a useful device, that must (and can) be ``added
by hand'' to conveniently describe charged sectors. In contrast, SLF, starting
from a perturbation theory supplemented by a new underlying ``principle of
string-independence'', first introduces free observables and charged fields on
the same footing, and then bases the distinction between the interacting
fields on the intrinsic difference in their causal localization properties.
This difference is particularly instrumental in the case of the charge
superselection rule.

\medskip

\noindent \textbf{Acknowledgments}

\noindent We are indebted to the authors of \cite{BCRV}, and
especially to D. Buchholz, to whom this paper is dedicated, for insisting in the importance of ``gauge
bridges'' for Gauss' Law in QED, and urging us to explain how we expect the
issue to be solved in the string-localized approach. We thank J.
Gracia-Bond\'ia and P. Duch for valuable comments. JM has received financial
support from the Brasilian research agency CNPq, and is also grateful to CAPES
and Finep.

\appendix

\section{Some background material}
We compile some pertinent facts about quantum fields as operator-valued
distributions. They may be obscured in the modern path integral treatments
which encode algebraic properties like Hilbert spaces and commutation relations
in a rather indirect way, and they fall outside the focus of
purely fibre-bundle views of gauge theory that usually do not talk
about operators at all. Along the way, we explain some arguments used
in the main text.

\subsection{Vector potentials and indefinite metric}
\label{a:IM}
Adding a gauge-fixing term $-\frac\lambda2(\partial A)^2$ to the QED Lagrangean results in the equations of motion
\bea{A1}
\square A^\nu = (1-\lambda)\partial^\nu (\partial A) -q\,\ol\psi \gamma^\nu \psi,
\eea
where $-q\,\ol\psi \gamma^\nu \psi=j^\nu_{\rm elm}$ is the
electromagnetic current. The field strength $F_{\mu\nu}=\partial_\mu A_\nu-\partial_\nu A_\mu$
trivially satisfies the homogeneous Maxwell equation
\bea{A2}
\partial_\mu F_{\nu\kappa} + \partial_\nu F_{\kappa\mu} + \partial_\kappa F_{\mu\nu} =0
\eea
and, as a consequence of \eref{A1}, the inhomogeneous equation
\bea{A3}
\partial_\mu F^{\mu\nu}= j^\nu_{\rm fict}+j^\nu_{\rm elm},
\eea
where the ``fictitious current'' $j^\nu_{\rm fict}=-\lambda \partial^\nu(\partial A)$
on the right-hand side is present also without interaction ($q=0$,
hence $j_{\rm elm}^\nu=0$). It
prevents the validity of Gauss' Law, because the integral over the
fictitious charge density is a non-vanishing operator. (In the Lorenz gauge $\lambda=\infty$, where
$(\partial A)=0$ but $\square A_\nu\neq 0$, the right-hand side of
\eref{A1} is not defined, and the 
fictitious current turns out to be $j^\nu_{\rm fict}=\square A^\nu$ instead.)

The unique covariant two-point function of a free quantum field
satisfying \eref{A1} with $q=0$ is 
\bea{A4}
(\Omega, A_{\mu}(x)A_{\kappa}(y)\Omega) =
-\int\frac{d^4k}{(2\pi)^{3}}\, \theta(k^0) \big(\eta_{\mu\kappa}\delta(k^2) +\big(1-\lambda^{-1}\big)k_\mu k_\kappa \delta'(k^2)\big) e^{-ik(x-y)}.\qquad
\eea
The two-point function determines the scalar product of states $\vert
f\rangle=\int d^4x \, f^\mu(x)A_\mu(x)\Omega$:
\bea{A5}
\langle f\vert g\rangle = \int\frac{d^4k}{(2\pi)^{3}}\, \theta(k^0)
\Big[-\eta_{\mu\kappa}\overline{\hat f^\mu(k)}\hat g^\kappa(k) \cdot\delta(k^2) -
\big(1-\lambda^{-1}\big)\overline{(k\hat f(k))}(k\hat g(k))\cdot\delta'(k^2)\Big] .\qquad\eea
These are obviously indefinite (there are states of negative
norm-square) because of the presence of the Lorentz tensor
and of the derivative $\delta'(k^2)$ (the latter
being absent in the Feynman gauge $\lambda=1$). The
indefinite-inner-product space generated by 
this field from the vacuum is called a ``Krein space''.

With the inner product \eref{A5}, one finds four
linearly independent states for each momentum, as opposed to the two
physical photon polarization states. The unphysical states have to be
eliminated by the Gupta-Bleuler or BRST method, defining the
physical Hilbert space as a quotient space (a semi-definite subspace
modulo the null states). States
generated by the fictitious current have norm-square zero and are
eliminated; but the vector potential is not defined on the quotient
space.

\newpage

The resulting two-point function of the field strength
\bea{A6}
(\Omega, F_{\mu\nu}(x)F_{\kappa\lambda}(y)\Omega) =
\int\frac{d^4k}{(2\pi)^{3}}\, \delta(k^2) \, \theta(k^0) \, k_{[\mu}
\eta_{\nu][\kappa}k_{\lambda]}\,
 e^{-ik(x-y)}
\eea
gives rise to a positive definite inner product, hence a Hilbert
space. The subspace generated from 
the vacuum by the field strength is well-known to coincide with the
physical quotient space, and it is the Fock space over the sum of the irreducible helicity
$\pm1$ Wigner representations (corresponding to the physical photon states). 

In particular, the string-localized vector potential 
\bea{A7}
A_\mu(x,e) := \int_0^\infty ds\, F_{\mu\kappa}(x+se)e^\kappa
\eea
is defined on the physical Hilbert space. One easily computes, using
the homogeneous Maxwell equation \eref{A2}: 
$$
\partial_\mu A_\nu(x,e)-\partial_\nu A_\mu(x,e) = -\int_0^\infty ds\,
\partial_\kappa F_{\mu\nu}(x+se) e^\kappa = -\int_0^\infty ds\, \frac
d{ds} F_{\mu\nu}(x+se) = F_{\mu\nu}(x).$$

On the other hand, the field strength is also defined on the
indefinite Krein space where $A_\mu(x)$ is defined. On this space, the
same string-localized potential (\ref{A7}) decomposes as
\bea{A8}
A_\mu(x,e) = \int_0^\infty ds\, (\partial_\mu
A_\kappa(x+se)-\partial_\kappa A_\mu(x+se))e^\kappa = A_\mu(x)  +
\partial_\mu\phi(x,e)\eea
where the string-localized field $\phi(x,e):= \int_0^\infty ds\, A_\kappa(x+se)e^\kappa$ is called
``escort field''.
Only the sum of the two terms on the right-hand side descends to the
physical Hilbert space, where the two terms are not seperately
defined.

The hybrid approach of \sref{s:Hy} makes use of the decomposition \eref{A8} in
the Krein space, in order to gain insight on the coupling of
$A_\mu(x,e)$ to the Dirac field on the physical Hilbert space. 

\medskip

\noindent {\bf Distributional aspects.}
Strictly speaking, string-localized fields are distributions in both variables
$x$ and $e$. Thus, they should be smeared with a test function
$f(x,e)$ so as to produce well-defined operators. In the case of the
QED interaction density $A_\mu(x,e)j^\mu(x)$, we choose
$f(x,e)=g(x)h(e)$ and think of $h$ as an averaging with $\int
d\sigma(e)h(e)=1$. The principle of 
string-independence requires that in the adiabatic limit $g(x)\to q$,
the resulting interacting observables do not depend of the details
of the function $h(e)$. In the case of the exponentiated escort field, the formal
expression $e^{iq\phi(x,e)}$ is too singular to be considered without
smearing. Thus, we smear the exponent with $f(x)h(e)$. Because of the
``semi-perturbative'' nature of the hybrid approach, it is sufficient
to study the exponential of the {\em free} escort field where the coupling constant
appears in the exponent. Thus, we do not need a cutoff function $g(x)$
for the interaction density. Instead, we regard both $f$ and $h$ as
averagings and absorb the coupling constant into the function
$f$. This yields the conditions $\int d^4x f(x)=q$ and $\int d\sigma(e) h(e)=1$.

\medskip

\noindent {\bf Massive case.}
In the massive case ($A_\mu=$ Proca field), there is no problem with indefinite metric.
The same formulas \eref{A7} and
\eref{A8} define a  string-localized potential and its escort field, both on
the physical Hilbert space. The benefit of $A_\mu(x,e)$, as compared to the 
Proca field, is in this case the improved short-distance behaviour of the
two-point function and the associated propagator, turning massive QED
into a renormalizable theory \cite{S,MRS}. Interestingly, the massless
string-localized potential  
is a smooth limit of the massive string-localized potential \cite{MRS}, in marked
contrast to the point-localized case.

\vskip-1mm

\subsection{Causality and observables}
\label{a:CO}
Einstein Causality is the fundamental principle that two
(experimentally realizable) operations
on a system taking place at spacelike distance must not influence each other.
Quantum field theory implements this principle by the postulate that the
commutator of two observables localized at spacelike distance
vanishes. Notice that only operators of observable quantities
correspond to realizable operations of a system. 

This makes of course no statement about non-observable quantities, but
it implies that fields of half-integer spin that anti-commute at
spacelike distance, cannot be observable. Only quadratic expressions
such as currents can be observable.

The Spin-Statistics theorem is ofted said to state that quantum fields
of half-integer spin must anti-commute at spacelike distance, and
must commute with integer-spin fields. A closer look at the
proof reveals that the correct statement is rather that half-integer spin fields
cannot commute at spacelike distance, and the usual argument is only
conclusive if no other options besides ``commute'' or ``anti-commute'' are allowed.
This conclusion is in particular correct for free fields. 

In string-localized QFT, ``spacelike distance'' means that the entire strings
along which fields may be localized, are spacelike separated, cf.\
footnote \ref{f:string}.

Independent of the Spin-Statistics connection, charged fields in
gauge theories are not observable because they are not
gauge-invariant. Consequently, there is no apriori reason why 
they should obey any specific spacelike commutation properties at all,
as long as their observable currents commute among themselves and with
all other observables.

\smallskip

\noindent {\bf Gauss' Law, localization, and superselection sectors.}
The argument of [FPS] on QED shows
that, if the global Gauss Law holds, the interacting Dirac field
cannot commute like a point-localized field with the Maxwell field at
spacelike distance. It can do so in the Krein space where the Gauss Law
fails. The present work shows that in the
Hilbert space the Dirac field becomes string-localized by the coupling
to the vector potential, so  
that it still commutes with the Maxwell field in spacelike separation
from the string. Because the string extends to infinity, it does not
commute with the electric flux at infinity, and the conflict with
Gauss' Law is resolved. 

Because the electric field is observable, the limits
\bea{A9}
\lim_{r\to\infty} r^2 F^{0i}(0,r\vec e)
\eea
(``asymptotic flux densities'' localized on the ``sphere at spacelike
infinity'') commute with every 
local observable (but certainly not with string-localized quantities).
By Schur's Lemma, they are multiples of $\mathbf{1}$ in every
irreducible representation of the algebra of local observables. Different
values of these operators characterize inequivalent representations (superselection sectors).

\newpage

In  \sref{s:IP}, we determine their values
in charged states by computing the expectation
values $(\Psi,f^{0i}(r\vec e)\Psi)$. If the charged states $\Psi$ are
generated from the vacuum by interacting Dirac field operators, we find that
the asymptotic flux density distribution has the same shape as the
smearing function of the string direction used for the interaction. 

The ``GNS construction'' mentioned in \sref{s:Phot} assigns to a
state (a positive expectation value functional on an algebra) a Hilbert space with a
representation of that algebra in which the state is realized by a vector.
Even if the functional is approximated by vectors in some Hilbert
space, the resulting GNS Hilbert space may be unrelated to the Hilbert
space used for the approximation. This happens in 
the case at hand, where the photon cloud superselection sectors are
not realized in a Fock space.

\medskip

\noindent {\bf Gauss' Law and infra-particles.}
As stated in \sref{s:IP}, weak convergence of the operator $\lim_{\lambda\to\infty}\lambda^2
F_{\mu\nu}(\lambda x)$ for spacelike $x$ to a homogeneous function $a_{\mu\nu}(x)$
times the unit operator implies that charged particles cannot have a
sharp mass \cite{Bu2}, i.e., one-particle states cannot be
eigenvectors of the mass operator. The (simplified) 
argument proceeds by observing that for eigenvectors $\Psi$ one has
$$0=(\Psi,[P_\kappa P^\kappa, F_{\mu\nu}(x)]\Psi) =
2i(P^\kappa \Psi,\partial_\kappa F_{\mu\nu}(x)\Psi) -
(\Psi,\square F_{\mu\nu}(x)\Psi). $$
In the asymptotic limit \eref{amunu}, $F_{\mu\nu}$ can be replaced by
$a_{\mu\nu}$, and the second term decays more
rapidly than the first, so that $\partial_\kappa a_{\mu\nu}(x)\cdot
(P^\kappa \Psi,\Psi)$ must vanish. Because the scalar product here is
generically non-zero, one concludes $\partial_\kappa a_{\mu\nu}(x)=0$
which by homogeneity entails $a_{\mu\nu}(x)=0$. Hence, $\Psi$ can be
an eigenvector only if its total
charge is zero.


\begin{thebibliography}{99}
\bibitem{ADS}{\small A. Aste, M. D\"utsch, G. Scharf: \emph{Perturbative gauge
    invariance: the electroweak theory. II}, Ann.\ d.\ Physik 8 (1999) 389--404}


\bibitem{BN} {\small F. Bloch and A. Nordsieck, \emph{Note on the
      radiation field of the electron}, Phys.\ Rev.\ 52 (1937) 54--59
}

\bibitem {BLOT}{\small N.N. Bogoliubov, A.A. Logunov, A.I. Oksak, I.T.
Todorov:  General Principles of Quantum Field Theory, Kluwer, Dordrecht, 1990
}

\bibitem {BS}{\small N.N. Bogoliubov, D.V. Shirkov: Introduction to the
Theory  of Quantized Fields, Wiley, New York, 1959 }

\bibitem {Br}{\small R.A. Brandt: \emph{Field equations in quantum
electrodynamics}, Fortschr.\ Physik 18 (1970), 249--283 }

\bibitem {Bu1}{\small D. Buchholz: \emph{The physical state space of
      Quantum Electrodynamics}, Commun.\ Math.\ Phys.\ 85 (1982) 49--71 }

\bibitem {Bu2}{\small D. Buchholz: \emph{Gauss' Law and the infraparticle
problem}, Phys.\ Lett.\ B174 (1986) 331--334 }

\bibitem {BCRV}{\small D. Buchholz, F. Ciolli, G. Ruzzi, E. Vasselli:
    \emph{On  string-localized potentials and gauge fields},
    Lett.\ Math.\ Phys.\ 109 (2019) 2601--2610}

\bibitem {BF}{\small D. Buchholz, K. Fredenhagen: \emph{Locality and the
structure of particle states}, Commun.\ Math.\  Phys.\ 84 (1982) 1--54 }


\bibitem {CMV}{\small L.T. Cardoso, J. Mund, J. V\'arilly: \emph{String
chopping and time-ordered products of linear string-localized  quantum
fields}, Math.\ Phys.\ Anal.\ Geom.\ (2018) 21:3 }

\bibitem {Do}{\small J.D. Dollard: \emph{Adiabatic switching in the
Schr\"odinger theory of scattering}, J. Math.\ Phys.\ 7 (1966) 802--810 }

\bibitem {D}{\small P. Duch: \emph{Infrared problem in perturbative quantum
field theory}, arXiv:1906.00940 }

\bibitem{DF}{\small M. D\"utsch, K. Fredenhagen: \emph{Algebraic quantum field theory, perturbation theory,
and the loop expansion}, Commun.\ Math.\ Phys.\ 219 (2001) 5--30}

\bibitem {DW}{\small W. Dybalski, B. Wegener: \emph{Asymptotic charges, large
gauge transformations and inequivalence of different gauges in external
current QED}, arXiv:1907.06750 }

\bibitem{DS}{\small M. D\"utsch, G. Scharf: \emph{Perturbative gauge
    invariance: the electroweak theory}, Ann.\ d.\ Physik 8 (1999) 359--387}

\bibitem {FK}{\small L. Faddeev, P. Kulish: \emph{Asymptotic conditions and
infrared divergences in quantum electrodynamics}, Theor.\ Math.\ Phys.\ 4
(1970) 745--757 }

\bibitem {FPS}{\small R. Ferrari, L.E. Picasso, F. Strocchi: \emph{Some
remarks  on local operators in Quantum Electrodynamics},
Commun.\ Math.\ Phys.\ 35 (1974) 25--38 }

\bibitem{FP} {\small W. Pauli, M. Fierz: \emph{Zur Theorie der
        Emission langwelliger Lichtquanten}, Nuovo Cim.\ 15 (1938) 167-188}


\bibitem {FMS}{\small J. Fr\"ohlich, G. Morchio, F. Strocchi: \emph{Charged sectors and scattering states in Quantum Electrodynamics},
Ann.\ Phys.\ 119 (1979) 241--284}


\bibitem {GMV}{\small J.M. Gracia-Bond\'ia, J. Mund,  J.C. V\'arilly:
\emph{The chirality theorem}, Ann.\ H. Poinc.\ 19  (2018) 843--874 }

\bibitem {H}{\small R. Haag: Local Quantum Physics, 2nd ed., Springer, 1996 }

\bibitem {M}{\small S. Mandelstam: \emph{Quantum Electrodynamics without
potentials }, Ann.\ Phys.\ 19 (1962) 1--24 }

\bibitem {MS1}{\small G. Morchio, F. Strocchi: \emph{A non-perturbative
approach to the infrared problem in QED: Construction of charged states},
Nucl.\ Phys.\ B211 (1983) 471--508; \emph{Errata}, Nucl.\ Phys.\ BB232 (1984)
547 }

\bibitem {MS2}{\small G. Morchio, F. Strocchi: \emph{Infrared problem in  QED
and electric charge renormalization}, Ann.\ Phys.\ 168 (1986) 27--45 }

\bibitem {MRS}{\small J. Mund, K.-H. Rehren, B. Schroer: \emph{Relations
between  positivity, localization and degrees of freedom: The Weinberg-Witten
theorem and the van Dam-Veltman-Zakharov discontinuity}, Phys.\ Lett.\ B 773
(2017) 625--631 }

\bibitem {MS}{\small J. Mund, K.-H. Rehren, B. Schroer: work in progress }

\bibitem {MSY}{\small J. Mund, B. Schroer, J. Yngvason: \emph{String-localized
quantum fields and modular localization}, Commun.\ Math.\ Phys.\ 268 (2006)
621--672 }

\bibitem {O}{\small C.A. Orzalesi: \emph{Charges and generators of symmetry
transformations in Quantum Field Theory}, Rev.\ Mod.\ Phys.\ 42  (1970)
381--408 }

\bibitem {S63}{\small B. Schroer: \emph{Infrateilchen in der
Quantenfeldtheorie}, Fortschr.\ Physik 11 (1963) 1--32 }

\bibitem {S}{\small B. Schroer: \emph{An alternative to the gauge theoretic
setting}, Found.\ Phys.\ 41 (2011) 1543--1568 }

\bibitem {S2}{\small B. Schroer: \emph{The role of positivity and causality in
interactions involving  higher spin}, Nucl.\ Phys.\ B 941 (2019) 91--144 }

\bibitem {St}{\small O. Steinmann: \emph{Perturbative QED in terms of gauge
invariant fields}, Ann.\ Phys.\ 157 (1984) 232--254 }

\bibitem {St2}{\small O. Steinmann: Perturbative Quantum Electrodynamics
and Axiomatic Field Theory, Springer (2000) }

\bibitem{Str} {\small A. Strominger: Lectures on the Infrared
    Structure of Gravity and Gauge Theory, Princeton University Press (2018)}

\bibitem {Sw}{\small J.A. Swieca: \emph{Charge screening and mass spectrum},
Phys.\ Rev.\ D13 (1976) 312--314 }


\bibitem {YFS}{\small D.R. Yennie, S.C. Frautschi, H. Suura, \emph{The
infrared  divergence phenomena and high-energy processes}, Ann.\  Phys.\ 13
(1961) 379--452 }
\end{thebibliography}
\end{document}